\documentclass[conf]{new-aiaa}
\usepackage[utf8]{inputenc}

\usepackage{graphicx}
\usepackage{amsmath}
\usepackage[version=4]{mhchem}
\usepackage{siunitx}
\usepackage{float}
\usepackage{caption}
\usepackage{subcaption}
\usepackage{hyperref}
\usepackage{longtable,tabularx}
\title{Evaluating eVTOL Network Performance and Fleet Dynamics through Simulation-Based Analysis}

\author{Emin Burak Onat\footnote{Ph.D. Candidate, Department of Civil and Environmental Engineering, University of California, Berkeley, Berkeley, CA 94720}}
\affil{University of California, Berkeley, Berkeley, CA 94720}

\author{Vishwanath Bulusu\footnote{Aerospace Research Scientist, Crown Consulting Inc., Aviation Systems Division, NASA Ames Research Center, Moffett Field, CA 94035}}
\affil{Crown Consulting Inc. at NASA Ames Research Center, Moffett Field, CA 94035}

\author{Anjan Chakrabarty\footnote{Lead Aerospace Research Engineer, Supernal, 401 Kato Terrace, Fremont, CA 94539}}
\affil{Supernal, 401 Kato Terrace, Fremont, CA 94539}

\author{Mark Hansen\footnote{Professor, Department of Civil and Environmental Engineering, University of California, Berkeley, Berkeley, CA 94720}}
\author{Raja Sengupta\footnote{Professor, Department of Civil and Environmental Engineering, University of California, Berkeley, Berkeley, CA 94720}}
\author{Banavar Sridhar\footnote{Visiting Professor, Department of Civil and Environmental Engineering, University of California, Berkeley, Berkeley, CA 94720}}
\affil{University of California, Berkeley, Berkeley, CA 94720}

\begin{document}

\maketitle

\begin{abstract}
Urban Air Mobility (UAM) represents a promising solution for future transportation. In this study, we introduce VertiSim\footnote{VertiSim is a constantly evolving tool. It will soon be available as an open-source tool. For further information, papers and updates, please refer to \href{https://air.berkeley.edu}{https://air.berkeley.edu}}, an advanced event-driven simulator developed to evaluate e-VTOL transportation networks. Uniquely, VertiSim simultaneously models passenger, aircraft, and energy flows, reflecting the interrelated complexities of UAM systems. We utilized VertiSim to assess 19 operational scenarios serving a daily demand for 2,834 passengers with varying fleet sizes and vertiport distances. The study aims to support stakeholders in making informed decisions about fleet size, network design, and infrastructure development by understanding tradeoffs in passenger delay time, operational costs, and fleet utilization. Our simulations, guided by a heuristic dispatch and charge policy, indicate that fleet size significantly influences passenger delay and energy consumption within UAM networks. We find that increasing the fleet size can reduce average passenger delays, but this comes at the cost of higher operational expenses due to an increase in the number of repositioning flights. Additionally, our analysis highlights how vertiport distances impact fleet utilization: longer distances result in reduced total idle time and increased cruise and charge times, leading to more efficient fleet utilization but also longer passenger delays. These findings are important for UAM network planning, especially in balancing fleet size with vertiport capacity and operational costs. Simulator demo is available at: \url{https://tinyurl.com/vertisim-vis}
\end{abstract}

\section{Introduction}
The advent of Urban Air Mobility (UAM), featuring electric vertical takeoff and landing (e-VTOL) aircraft, presents an exciting new dimension to urban transportation. However, effective integration and optimization of these novel air networks bring unprecedented challenges, necessitating robust and advanced simulation tools. This paper introduces VertiSim as a network evaluation tool for UAM, an enhanced event-driven simulator designed to model and optimize e-VTOL transportation networks, addressing the inherent complexities through a holistic approach.

VertiSim distinguishes itself from legacy aviation simulators by simultaneously modeling three critical flows: passenger, aircraft, and energy. This integrated approach reflects the unique operational characteristics of e-VTOL networks where flight durations, passenger waiting times, and aircraft charging times  at vertiports exhibit comparable magnitudes. This parity in time scales signifies that any of these elements could emerge as a bottleneck in the system, necessitating a comprehensive understanding and detailed study of each flow for a holistic evaluation of system performance. VertiSim processes these interdependent flow into key metrics instrumental for effective network operation, such as aircraft utilization, average load factor, vertiport throughput, terminal area congestion (number of aircraft per unit time per unit area), utilization of resources, energy consumption for each flight phase, number of repositioning flights, and passenger waiting and trip times.

We use VertiSim to examine 19 e-VTOL operational scenarios involving different fleet sizes and vertiport distances. The paper elaborates the architectural design and capabilities of VertiSim, provides an in-depth analysis of the tested scenarios, and discusses the implications of the findings for future UAM network operations and fleet management.

\section{Literature Review}
UAM network simulation is a dynamic and rapidly growing field of research, attracting interest from a multitude of disciplines and sparking a myriad of novel approaches and methodologies.

Kohlman and Patterson \cite{Kohlman} focus on energy-related constraints, such as eVTOL aircraft battery life and the number of charging stations required for ground infrastructure. They provide a system-level model of the number of vehicles needed and the time these vehicles may have to loiter before landing. However, there is a need to capture network energy consumption accurately. Critical variables like occupancy, speed, and aircraft aerodynamics do not influence energy consumption in their model, and charging times are oversimplified. Furthermore, their model lacks passenger dynamics.

In a study comparing air taxis and traditional taxicabs \cite{airtaxi_taxicab}, two separate tools were coupled - Gridcity, a UAM traffic scenario generator, and NDMap, a conflict detection and resolution software. Despite yielding comparative data on route lengths and travel time, this approach is not an agent-based simulation, thus unable to detect conflicts or performance metrics such as holding times at vertiports due to infrastructure limitations.

The development of SimUAM \cite{simUAM} marked an important step towards understanding UAM's value proposition for metropolitan areas. It is a multi-modal toolchain that integrates ground traffic microsimulation, ground-air interfacing at the vertiport, and aerial microsimulation. However, the tool's design, which loosely couples multiple components, tends to lose track of congestion in the transition between simulation tools, as the output of one tool becomes the input for another.

A discrete-event simulation (DES) approach was employed by \citep{RAJENDRAN} for the study of emerging air taxi network operations. The authors utilized SIMIO® simulation software to develop a DES model of air taxi network operations, primarily focusing on determining the required number of air taxis to fulfill demand. While their approach offers a comprehensive view of network operations, the model does not incorporate charging stations or the energy required to execute the missions.

Li et al. \cite{Kochenderfer} presents a comprehensive examination of UAM network design decision variables, covering both strategic and tactical aspects. The study delves into long-term strategic variables, such as infrastructure and network design, as well as short-term strategic variables like route design and fleet size. It also explores tactical variables, including fleet management strategies and operation rules. This approach provides an integrated perspective on optimizing the UAM ecosystem. Furthermore, the study incorporates a macroscopic simulation to model UAM operations and estimate capacity and throughput, lending a high-level view of overall system performance. While this work is effective for analyzing and understanding the overall performance of the network, it lacks the detailed insights that a microscopic simulation would offer. Additionally, the model does not incorporate a charging policy, omitting a crucial aspect of eVTOL operations.

Evaluating design decision variables like demand patterns, network topology, vehicle type, and operational parameters interdependently is of paramount importance for designing efficient UAM networks. In response to a clear need for a more holistic, flexible, and precise simulation tool in UAM operations, we developed VertiSim. As an event-driven simulator, it is purpose-built to model e-VTOL transportation networks by capturing the complex interdependencies of passenger, aircraft, and energy flows, a capability not found in prior models. Uniquely, VertiSim incorporates the critical consideration of aircraft battery usage, operational range, and charging times, and also allows detailed modeling of the charging process. By modeling passengers and aircraft concurrently, VertiSim offers a more detailed, accurate, and comprehensive understanding of network dynamics, standing out from other studies in the field. As the only open-source tool that caters to this need, VertiSim can be an important resource for UAM network simulation. The goal is to offer an accessible and adaptable platform that encourages further exploration and refinement in this rapidly developing field.

\section{VertiSim: UAM Network Performance Evaluation Tool}
In a preceding work, we introduced VertiSim, an event-driven simulator, aimed at modeling a single vertiport operations \cite{vertisim}. The paper provided a comprehensive exploration of VertiSim's software architecture and modules. Building on this foundational knowledge, we present significant enhancements to VertiSim in this paper, with a new ability to simulate a full network of vertiports.

\begin{figure}[!htb]
  \centering
  \includegraphics[width=0.6\textwidth]{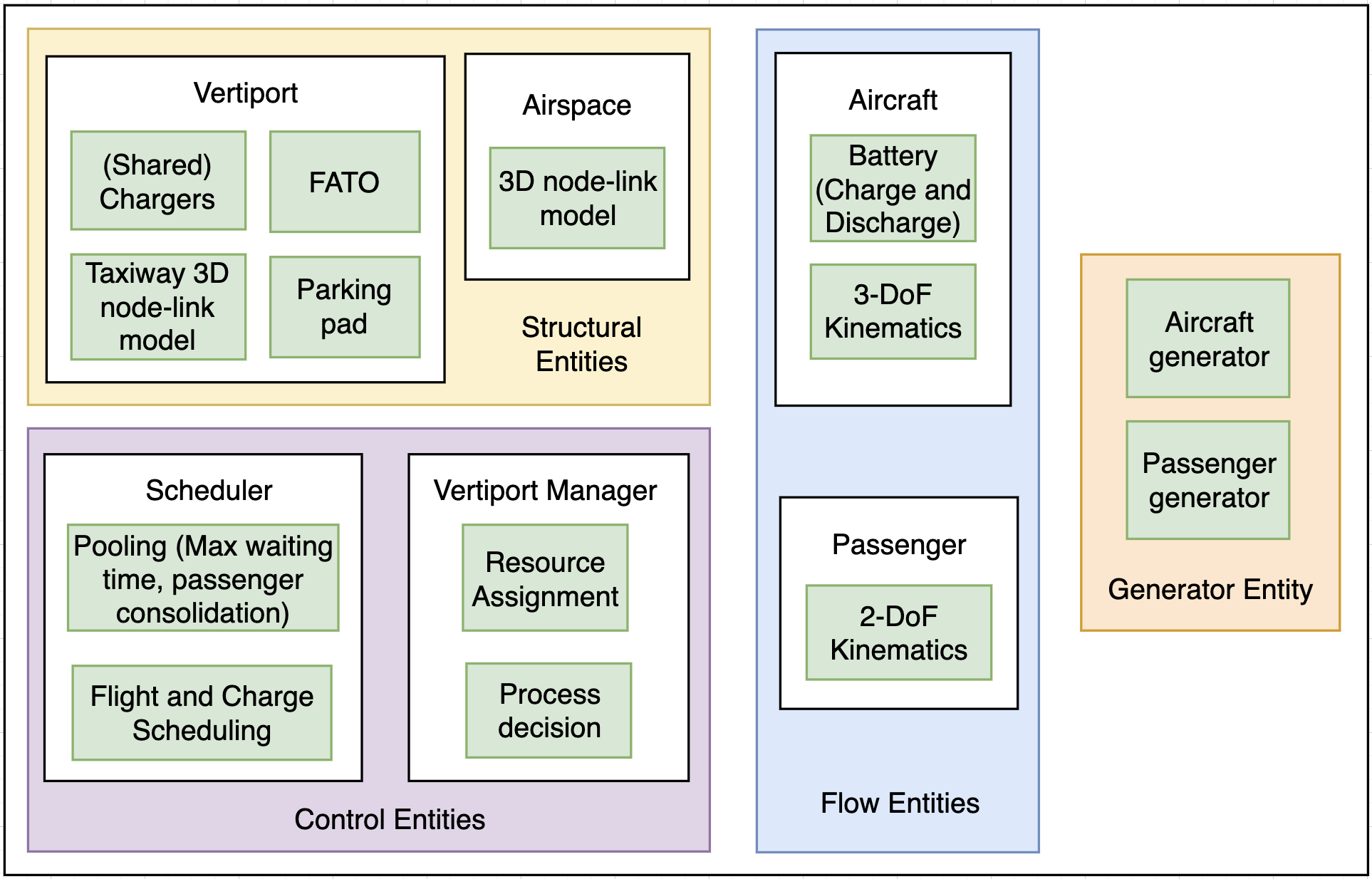}
  \caption{Building blocks of VertiSim}\label{fig:building_blocks}
\end{figure}

The initial design of VertiSim was centered around a building-block approach to model the vertiport simulator, inspired by the concept detailed in \cite{airportterminalmodeling}. This approach allowed us to construct comprehensive and customizable vertiport models to accommodate a variety of design and operational requirements. The core building blocks comprise \textit{structural} entities, \textit{flow} entities, \textit{generator} entities, and \textit{control} entities shown in figure \ref{fig:building_blocks}. Structural entities form the static layout of the simulation environment with elements like queues and servers. Flow entities such as aircraft, passengers, passenger groups, and energy interact with and move within these structural entities. Generator entities produce flow entities based on their schedule, while control entities handle tasks such as assigning TLOF (Touch-down and Lift-Off area) for arrival and departure, allocating parking pads, determining service priorities, dispatching flights, initiating charging and routing.

To further enhance the capability of VertiSim, we have introduced new features in this work. An airspace module has been added, enabling VertiSim to model the eVTOL aircraft flight paths between various vertiports within a network. We have implemented a speed optimization module to determine the energy-optimal speed for each flight phase and implement strategic dispatch and charging policies. The aircraft module has also been extended to include a detailed model for energy consumption across different flight phases. With these advancements, VertiSim now offers a comprehensive, detailed, and realistic simulation of a vertiport network.

\subsection{Software Architecture}

\begin{figure}[!htb]
  \centering
  \includegraphics[width=0.8\textwidth]{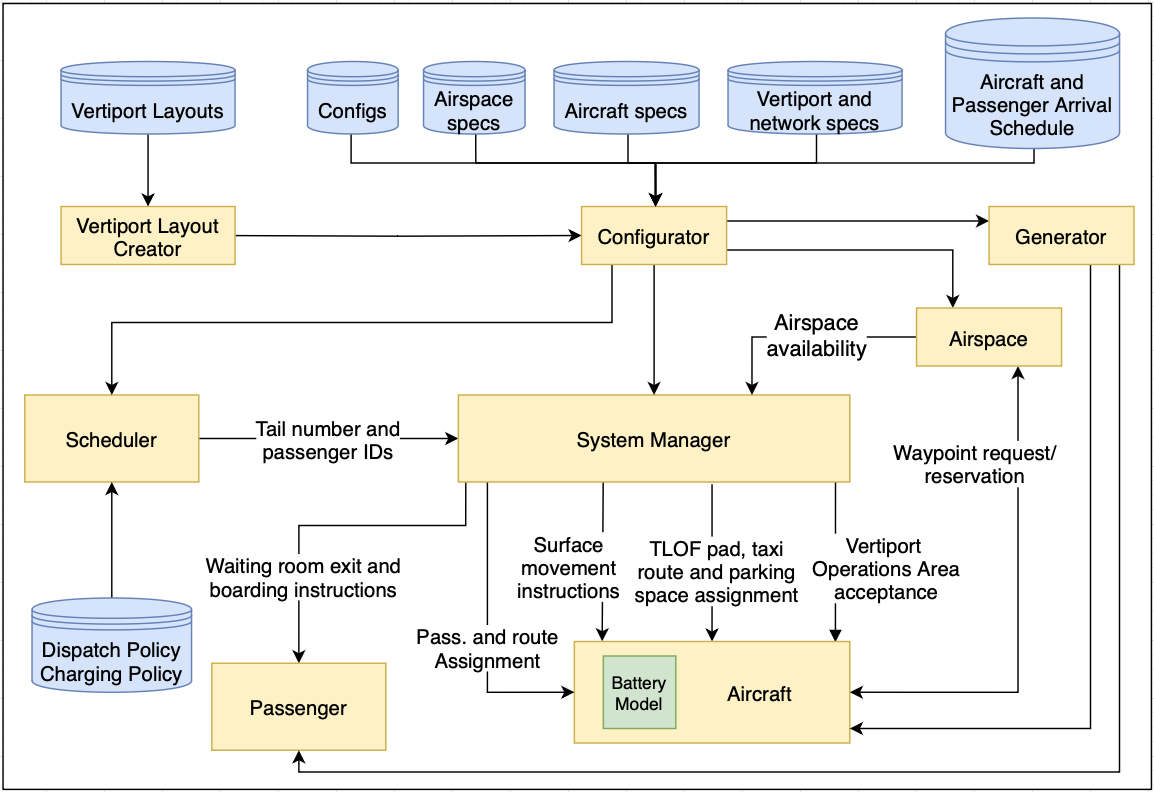}
  \caption{VertiSim Software Architecture}\label{fig:VertiSim_software_arch}
\end{figure}

VertiSim employs a graph-theoretic methodology to build structural entities based on the input vertiport layout. The simulation environment is conceptualized as a network of nodes and links, $G(V, E)$ in which $V = \{v_i\}$ is its vertex set and $E = \{(v_i, v_j)\}$ its edge set. Node positions are sourced from the input data. In VertiSim, every node and link is treated as a resource with a user-specified capacity. The software architecture of VertiSim comprises nine main modules (\textit{Passenger}, \textit{Aircraft}, \textit{System Manager}, \textit{Scheduler}, \textit{Airspace}, \textit{Vertiport Layout Creator}, \textit{Configurator}, \textit{Generator} and \textit{Battery}) to simulate vertiport network processes as depicted in figure \ref{fig:VertiSim_software_arch}.

\subsubsection{Main Modules}
The \textit{Vertiport Layout Generator} processes the input vertiport layouts to create a node-link graph model, depicting physical structures like TLOFs, spots, parking pads, and more as nodes, and taxiways and ramps as links. It also handles the allocation of chargers, defined by quantity, maximum charge power, and charging efficiency.

The \textit{Configurator} module refines the structural entities generated by the \textit{Vertiport Layout Generator}, providing specific functions like assigning server roles to TLOF nodes for handling queues. It can also add and configure elements like chargers and security checkpoints with distinct service rates.

Airspace construction, facilitated by the \textit{Airspace} module, also uses the node-link graph approach. It represents the airspace as a 3D graph of nodes and edges which eVTOLs can follow. It offers flexibility for various scenario modeling.

An \textit{Aircraft} is defined by state variables, including its current location, forward velocity, vertical velocity, state of charge (SoC in \%), the IDs of its origin and destination vertiports, the serviced time at its current location, a list of passengers onboard, active process (idle, charging, descent, descent transition, hover descent, hover climb, climb transition, climb, cruise, holding, pushback, taxi) and its priority level. These variables allow the simulator to accurately track and model each aircraft's state and behavior within the vertiport network.

\textit{Passengers} appear at vertiport entrances or on arriving aircraft. After check-in, they wait for flight assignment by the Scheduler, then proceed to the gate under the System Manager's direction. Arriving passengers exit the vertiport after disembarking.

The simulator accepts \textit{Aircraft} and \textit{Passenger} agent arrival inputs for network simulation. The \textit{Aircraft} input defines its initial state, while \textit{Passenger} input establishes the origin, destination, and arrival time.  The \textit{Generator} then creates these agents according to the \textit{Configurator} when the simulation clock reaches their designated arrival time. 

The \textit{System Manager} oversees vertiport operations, managing resources like TLOF pads, parking pads, airspace fixes, chargers, and security checkpoints as queues and servers with limited capacities. The \textit{System Manager} uses user-defined algorithms such as Dijkstra's shortest path to determine the most efficient route, and it ensures that the route is conflict-free, taking into account both upstream and downstream requirements. This conflict-free routing allows for simultaneous operations while ensuring no overtaking, no separation violation, or head-on conflict occurs on the surface. Furthermore, to avoid take-off and landing conflicts, the \textit{System Manager} ensures only one type of operation occurs at a TLOF at any one time. If multiple TLOFs are present, TLOFs can operate simultaneously if there is a 200ft separation between the TLOFs \cite{heliportdesign}. This separation limit can be set by the user.

The \textit{Scheduler} maintains a record of the number of \textit{Passengers} in each waiting room and dispatches flights according to the set rules. A flight can be initiated either when the number of passengers in the waiting room is sufficient to fill the aircraft's passenger capacity, or when a passenger's wait time surpasses a predetermined threshold.

\section{Simulation-Driven Analysis for UAM Network Performance}
\subsection{Vertiport Network and Infrastructure}
\begin{figure}[!ht]
  \centering
  \includegraphics[width=0.35\textwidth]{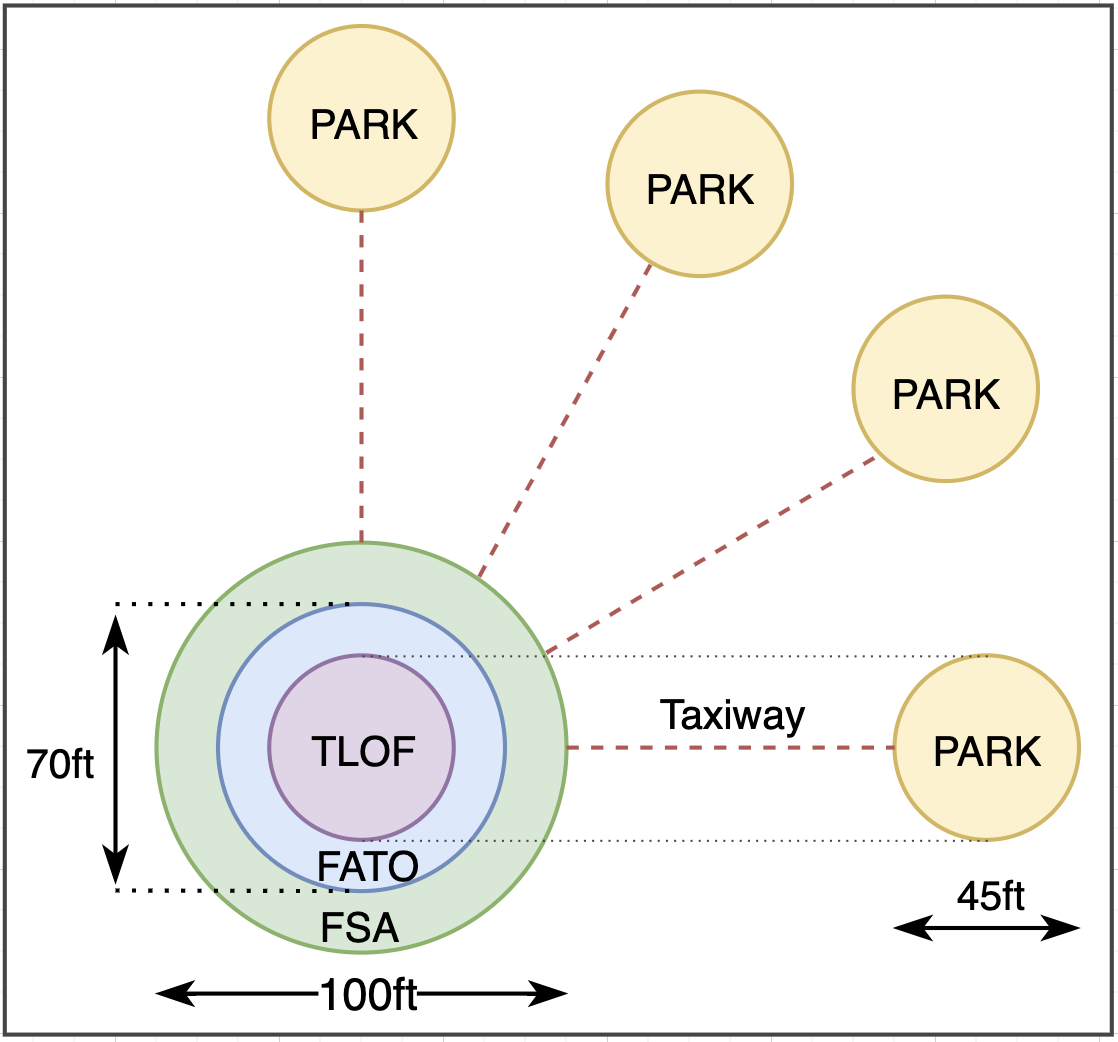}
  \caption{Clover type vertiport layout (FSA: Final Safety Area, FATO: Final Approach and Takeoff Area)}\label{fig:clover_layout}
\end{figure}

\begin{figure}[!ht]
  \centering
  \includegraphics[width=0.7\textwidth]{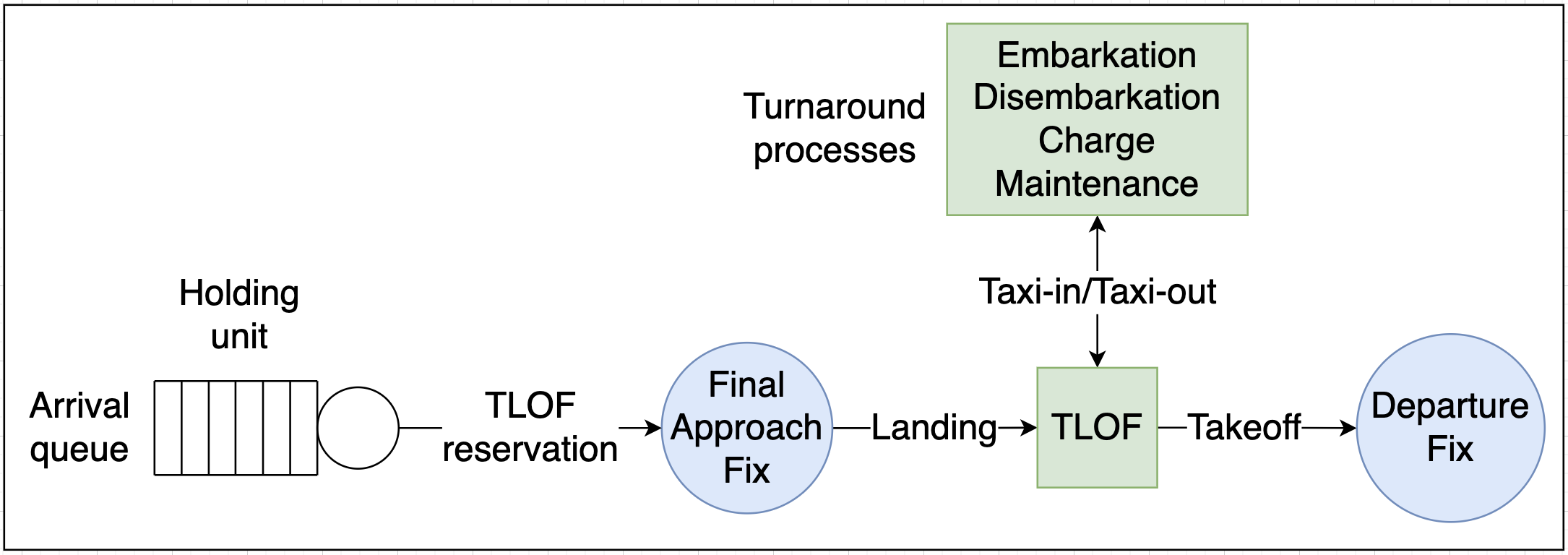}
  \caption{Vertiport operations}\label{fig:vert_ops}
\end{figure}

We selected networks of two vertiports with distances of 12, 24, and 36 miles, reflecting the range of urban trip lengths targeted by UAM services. Each vertiport consists of a TLOF, taxiways, and parking pads. Figure \ref{fig:clover_layout} presents a simplistic clover layout based on the FAA's Engineering Brief on vertiports \cite{faa_brief} and Zelinki's surface element spacing requirements \cite{ShannonVertiportTopology}, which is used in this study. Figure \ref{fig:vert_ops} depicts the conceptual framework for vertiport operations that we utilized in our study.

The vertiport operations start upon the arrival of an aircraft at a vertiport's holding unit. The holding unit is akin to the holding airspace around a vertiport. The aircraft then enters the landing queue. It must first secure a parking pad, move to the final approach fix and then reserve a TLOF. If no parking pads are available, the aircraft requried to remain in the holding unit until the \textit{System Manager} reserves one. However, in rare instances, the holding unit might also be at full capacity, due to the configuration of the system or the dispatch algorithm. In such cases, the aircraft will queue at the airspace waypoints. Once the pad is secured, a reservation for a TLOF is needed for landing. The descent transition phase begins when the \textit{System Manager} approves the TLOF reservation, followed by a hover descent for landing.

An aircraft may arrive carrying passengers and require charging for its next mission, or it may arrive empty and depart after charging or without charging for fleet repositioning. Passenger and aircraft agent interaction begins when the aircraft is ready for boarding, which occurs a $t_{boarding}$ time before charging completion. This simultaneous process ensures no delay in the aircraft turnaround time. When ready for departure, the aircraft seeks permission for TLOF and taxiing from the \textit{System Manager}. Once granted, taxiing can begin. Departure separation is achieved via the reservation of the departure fix.

The theoretical model for comparing surface capacity with TLOF capacity to identify whether a vertiport's throughput capacity is constrained by its parking or TLOFs was first proposed by \citep{guerreiro_scheduler}. This model was later expanded by \citep{vertisim}, as illustrated in equations \ref{eqn:surf_capacity} through \ref{eqn:vertiport_capacity}.

\begin{equation}\label{eqn:surf_capacity}
    C_{surf} = N_{park} \cdot \frac{t_{window}}{t_{arr}+t_{taxi-in}+t_{turnaround}+t_{taxi-out}+t_{dep}}
\end{equation}

\begin{equation}\label{eqn:TLOF_capacity}
    C_{TLOF} = 2 \cdot N_{TLOF} \cdot \frac{t_{window}}{t_{arr} + t_{taxi-out} + t_{dep}}
\end{equation}

\begin{equation}\label{eqn:vertiport_capacity}
    C_{vertiport} = min(2 \cdot C_{surf}, C_{TLOF})
\end{equation}

In these equations, $C_{surf}$ represents the number of surface operations that can be processed within a given time window, denoted as $t_{window}$. $C_{TLOF}$ stands for the number of balanced (alternating arrivals and departures) TLOF operations that can be processed in the same time window. Lastly, $N_{park}$ refers to the number of accessible parking nodes. Considering that we allot 60 seconds each for an arrival ($t_{arr}$) and a departure ($t_{dep}$), and 30 seconds each for taxi-in ($t_{taxi-in}$) and taxi-out ($t_{taxi-out}$), the TLOF capacity is calculated to be 24 arrivals and 24 departures per hour.

\subsection{Passenger Arrival Process}
Designing an accurate UAM network design involves the generation of a realistic passenger arrival process. To address this, we utilize data from the Performance Measurement System (PeMS) \cite{pems} collected at San Francisco Bay Bridge, adopting a granularity of five minutes. The data is sourced from the eastbound (VDS 402827) and westbound (VDS 402815) segments. The collection points are strategically located over a mile away from any ramps to ensure that the ramp traffic had a minimal impact on the data. To ensure that the data is representative of typical traffic patterns, we average two months of weekday data from April 2022 to May 2022. We note that demand is not balanced throughout a 24-hour period. There are peak periods of demand, reflecting rush hours, and periods of lower demand. 

To accommodate these variations in demand while maintaining operational feasibility, we scale down the hourly volumes to align with the vertiport throughput capacity, as determined by equation \ref{eqn:vertiport_capacity}. Subsequently, we employ a Poisson process to create passenger arrival times for each hour of the day. We assume that vertiports would have sufficient footprint to accommodate the required parking pads and would thus be limited only by TLOF capacity, thereby ensuring the simulation remains within system limits. The resulting dataset represents a daily flow of 1417 passengers per direction.

\begin{figure}[!ht]
  \centering
  \includegraphics[width=0.6\textwidth]{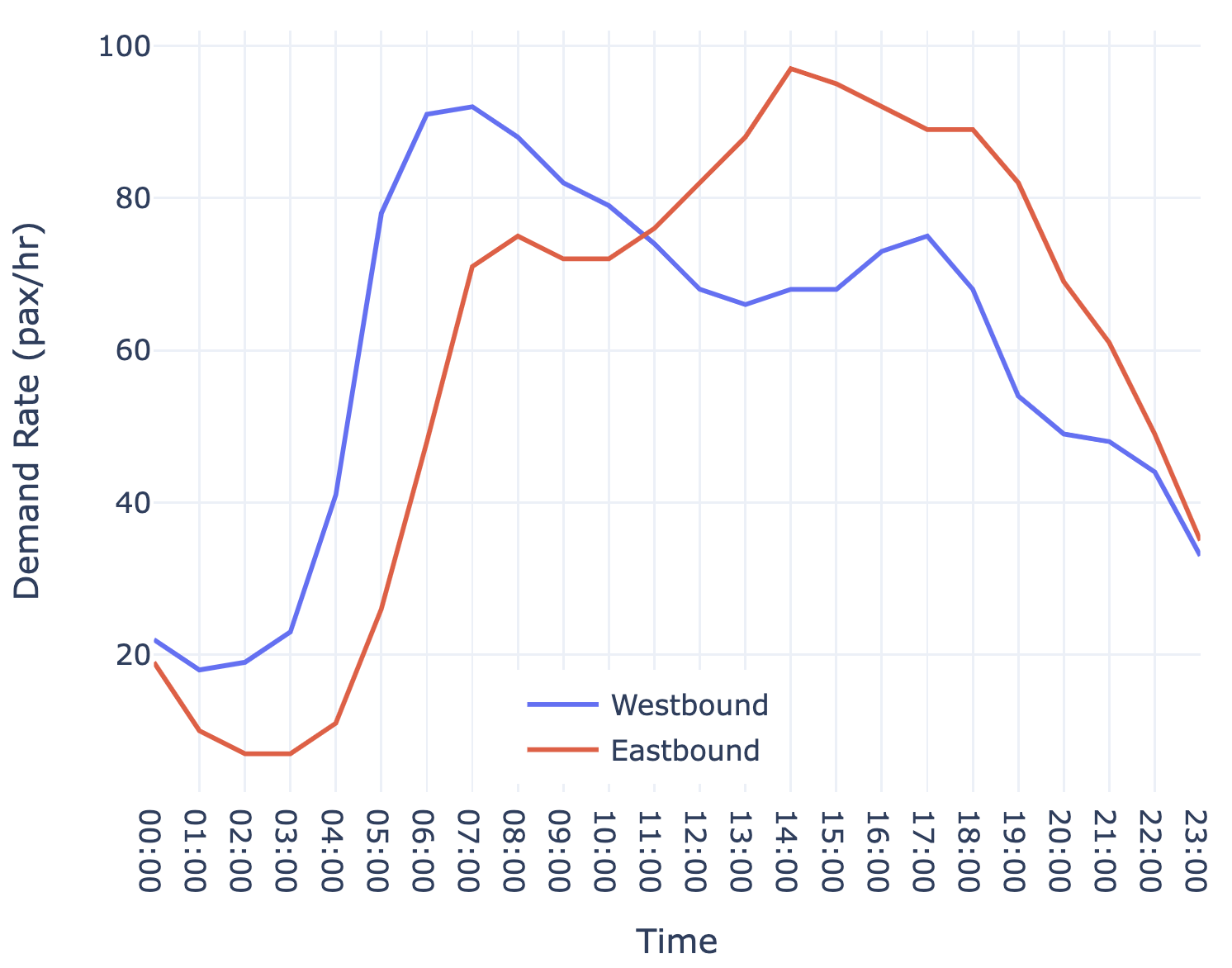}
  \caption{Scaled-down 24-hour demand between Eastbound and Westbound at the San Francisco Bay Bridge}\label{fig:demand1hr}
\end{figure}

\subsection{Estimated Departures}
\begin{figure}[!ht]
  \centering
  \includegraphics[width=0.55\textwidth]{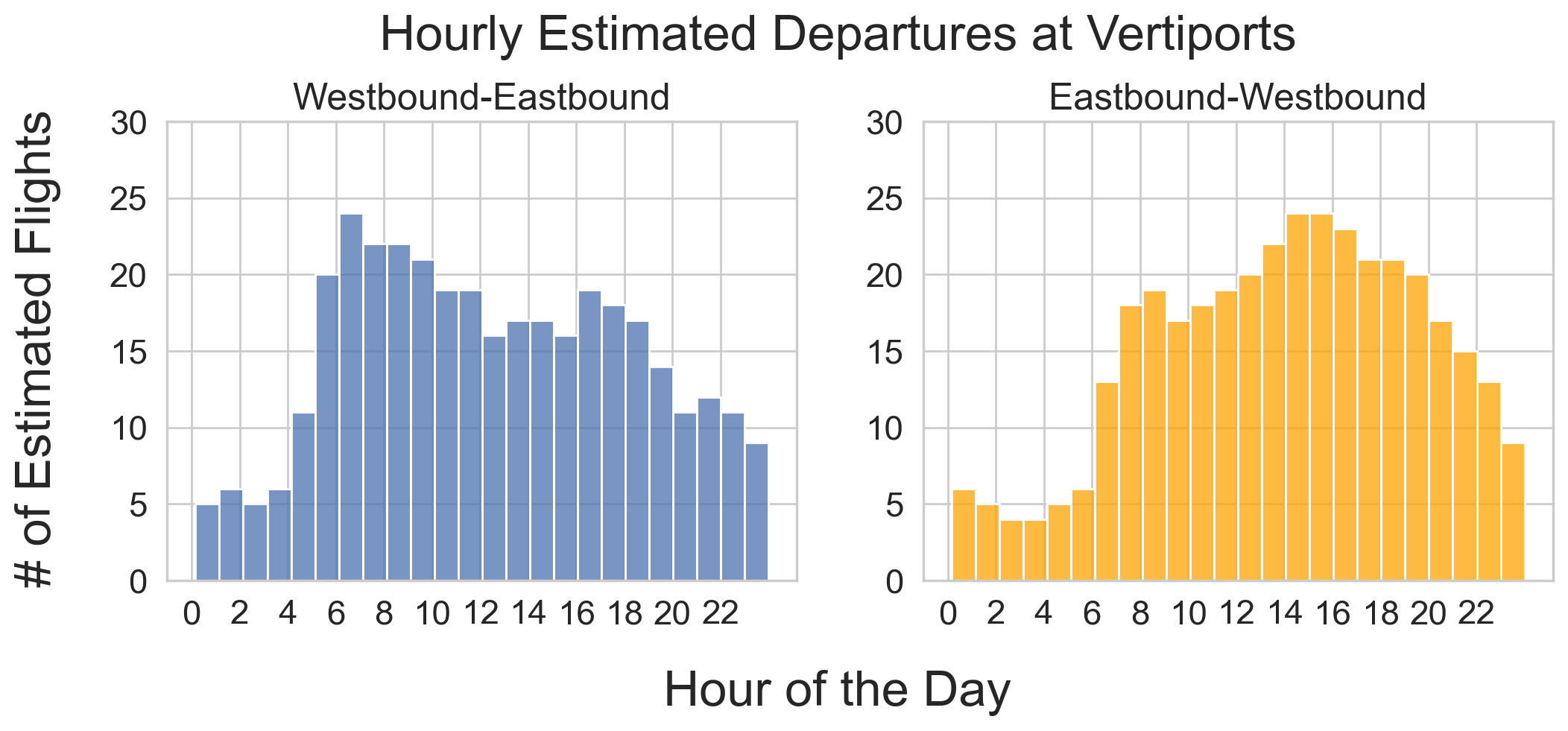}
  \caption{Hourly estimated departures at each vertiport}\label{fig:hourly_flight_schedule}
\end{figure}

In establishing the preliminary sizing of each vertiport within the network, we estimate the expected hourly flight departure volumes. We then implement a deterministic dispatch policy, dictating that an eVTOL flight is dispatched either when the vehicle reaches its full passenger capacity or when a passenger's wait time exceeds a pre-defined threshold. With a presumed eVTOL capacity of four passengers and a set threshold wait time of 10 minutes, we generate an illustrative representation of the day-long flight demand in both directions, presented on an hourly basis, as depicted in figure \ref{fig:hourly_flight_schedule}. We observe that the demand never surpasses the TLOF capacity of 24, indicating that a second TLOF is not necessary for the current operation scale. However, it does reach this capacity during certain peak hours of the day. Lastly, the dispatch policy produces 357 flights from Westbound to Eastbound and 363 flights from Eastbound to Westbound.

\subsection{Aircraft Model}
We employ the aircraft model for Joby S4, a tilt-rotor aircraft. Tilt-rotor aircraft can vertically take off and land like helicopters, but once airborne, they can convert to a more aerodynamically efficient fixed-wing configuration for forward flight \cite{tiltrotor}. The choice of the Joby S4 is primarily based on its status as an existing and well-accepted eVTOL aircraft, and on the availability of data in the literature \cite{joby_nyc, evtolbattery}. The mathematical modeling of our aircraft's operation is derived from equations specified in \cite{evtolbattery}. The model takes into account the power needed for the various flight stages, each possessing distinctive energy needs based on the physics of flight and the aircraft's characteristics. The power calculations for three primary flight stages - takeoff and landing, climb and descent, and cruise - are detailed as follows:

\textbf{Takeoff and Landing Power}:
\begin{equation}
P_{fixed-wing} = [\frac{f W}{FoM} \sqrt{\frac{f W / A}{2 \rho}} + \frac{W V_{climb_v}}{2}]/ \eta_{hover}
\end{equation}

\textbf{Climb and Descent Power}:
\begin{equation}
P_{fixed-wing} = [W V_v + \frac{1}{2}\rho V^3 S C_{D_0} + \frac{K W^2}{\frac{1}{2} \rho V S}] / (\eta_{climb})
\end{equation}

\textbf{Cruise Power}:
\begin{equation}
P_{fixed-wing} = [W V_v + \frac{W V}{[L/D]}] / (\eta_{cruise})
\end{equation}

Where $K$ represents the lift-induced drag coefficient, given by $K = 1/(4*C_{D_0}*(L/D)_{max}^2)$. The definition of the parameters are provided in table \ref{tab:parameters}.

The aircraft operation is modeled to utilize the minimum power for the climb, climb transition, descent, and descent transition phases. The cruise phase is operated for maximum range, resulting in segment-specific forward velocity, which is around 150 mph. For further insights into these computational details, refer to \cite{evtolbattery}.

During the simulation for each flight, the passenger weight is deducted from the total aircraft weight according to the number of unoccupied seats. The lift-to-drag ratio ($L/D$) is also specific to each flight segment. During the cruise, $L/D$ is set to 18, whereas for the climb and descent phases, it is considered to be 15.6. The air density, denoted by $\rho$, is dependent on both the atmospheric conditions and the altitude of the aircraft. At the ground level, we assume standard atmosphere conditions.

\begin{table}[!h] 
\caption{Aircraft model parameters \cite{evtolbattery}}
\centering
\label{tab:parameters}
\begin{tabular}{|c|c||c|c|}
\hline
Parameter                                                                                           & Value               & Parameter                                                                            & Value \\ \hline
Max Takeoff Weight (MTOM), W                                                                                  & 2182 {[}kg{]}       & Maximum lift coefficient, $C_L$                                                     & 1.5   \\ \hline
\begin{tabular}[c]{@{}c@{}}Correction factor for\\  interference from the fuselage, f\end{tabular} & 1.03                & \begin{tabular}[c]{@{}c@{}}Maximum lift to drag\\ ratio, $(L/D)_{max}$\end{tabular} & 15.3 - 18    \\ \hline
Disk load                                                                                           & 45.9 {[}kg/$m^2${]} & $\eta_{hover}$                                                                       & 0.85  \\ \hline
Wing area, S                                                                                           & 13 {[}$m^2${]}      & $\eta_{climb}$                                                                       & 0.85  \\ \hline
Figure of Merit, FoM                                                                              & 0.8                 & $\eta_{descent}$                                                                     & 0.85  \\ \hline
\begin{tabular}[c]{@{}c@{}}Zero lift drag \\ coefficient, $C_{D0}$\end{tabular}                     & 0.015               & $\eta_{cruise}$                                                                      & 0.90  \\ \hline
Passenger weight                                                                                    & 100 {[}kg{]}        & Atmosphere Conditions                                                                & Standard  \\ \hline
\end{tabular}
\end{table}

\subsection{Network Route and Flight Profile} 
In the formation of the network route, the automated airspace creator feature in VertiSim is applied. This tool is tasked with the generation of evenly-spaced nodes, set 1 mile apart. This specific distance is selected to ensure a separation of 1 mile between cruising aircraft. The nodes are placed between the initial cruise level node of the origin vertiport and the holding unit of the destination vertiport, creating a direct flight path. Node spacing can be readily adjusted to incorporate other separation standards. 

We employ the flight profile used in \cite{evtolbattery}. The flight profile comprises hover climb, climb transition, climb, cruise, descent, descent transition, and hover descent.

\begin{figure}[!ht]
  \centering
  \includegraphics[width=0.7\textwidth]{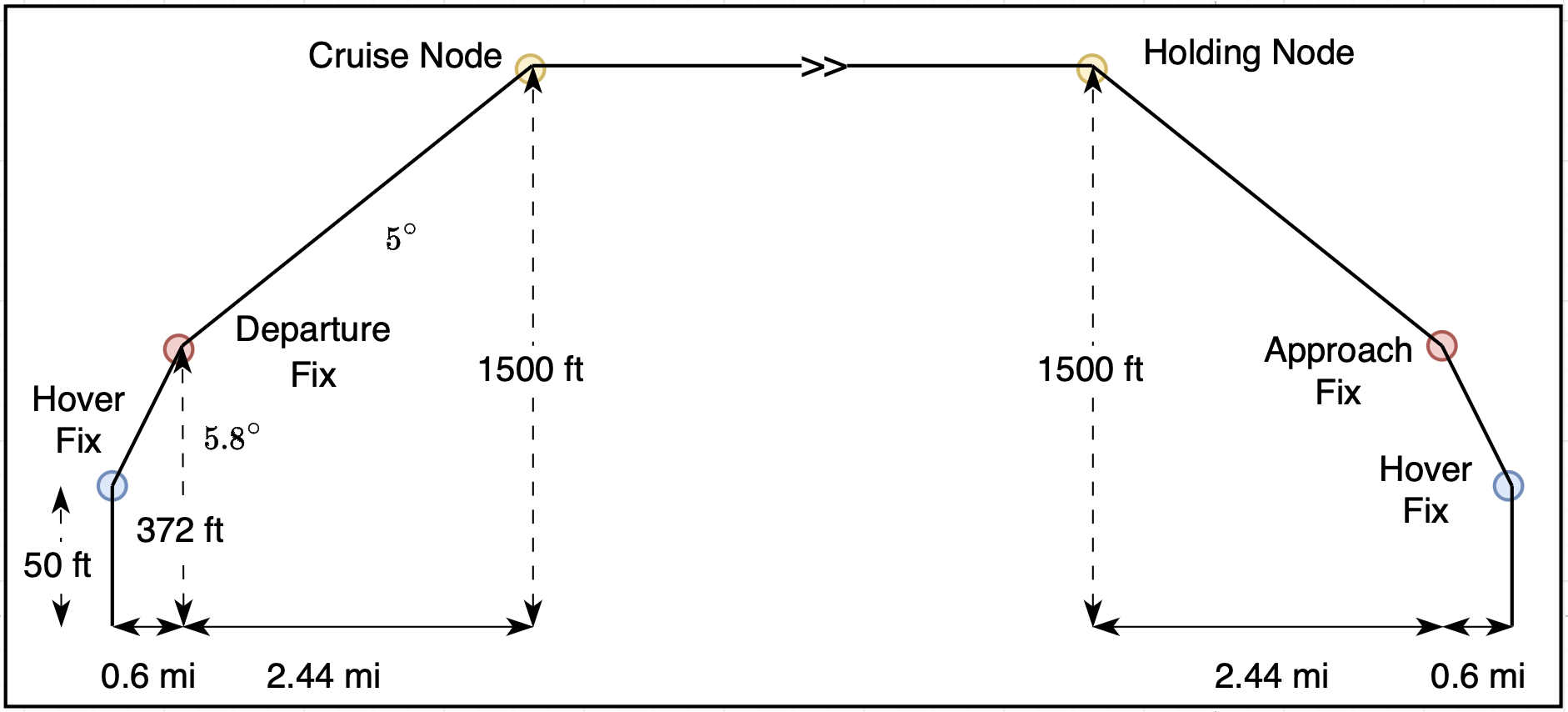}
  \caption{Flight profile}\label{fig:flight_profile}
\end{figure}

\subsection{Aircraft Battery Sizing}
Our battery sizing methodology is based on the assumption that each eVTOL aircraft must retain at least a 20\% State of Charge (SoC) reserve following each mission. This means that 80\% of the battery's SoC should be adequate to complete the longest possible flight that is pledged by the manufacturer, even under maximum passenger load. Sripad and Viswanathan \cite{evtolbattery} built a model to compute the energy consumption (Wh/pax-mi) for Heaviside, Joby S4, Lilium Jet, Alia, and Archer Maker eVTOLs by modeling aircraft physics and simulating the take-off, hover up, climb transition, cruise, descent transition, hover down, and landing processes of these models. We use their energy consumption model to determine the energy consumption at this maximum range, with the added 20\% SoC reserve factored in. The total energy thus calculated serves as our benchmark for battery capacity.

Our analysis found that the Archer Maker is equipped with a 75 kWh battery, which precisely corresponds with the official specifications released by Archer \cite{archer}. Moreover, 5-seater Joby S4, an eVTOL model used in our study, was computed to have a battery capacity of 160 kWh.

\subsection{Charging Model}

\begin{figure}[!ht]
  \centering
  \includegraphics[width=0.5\textwidth]{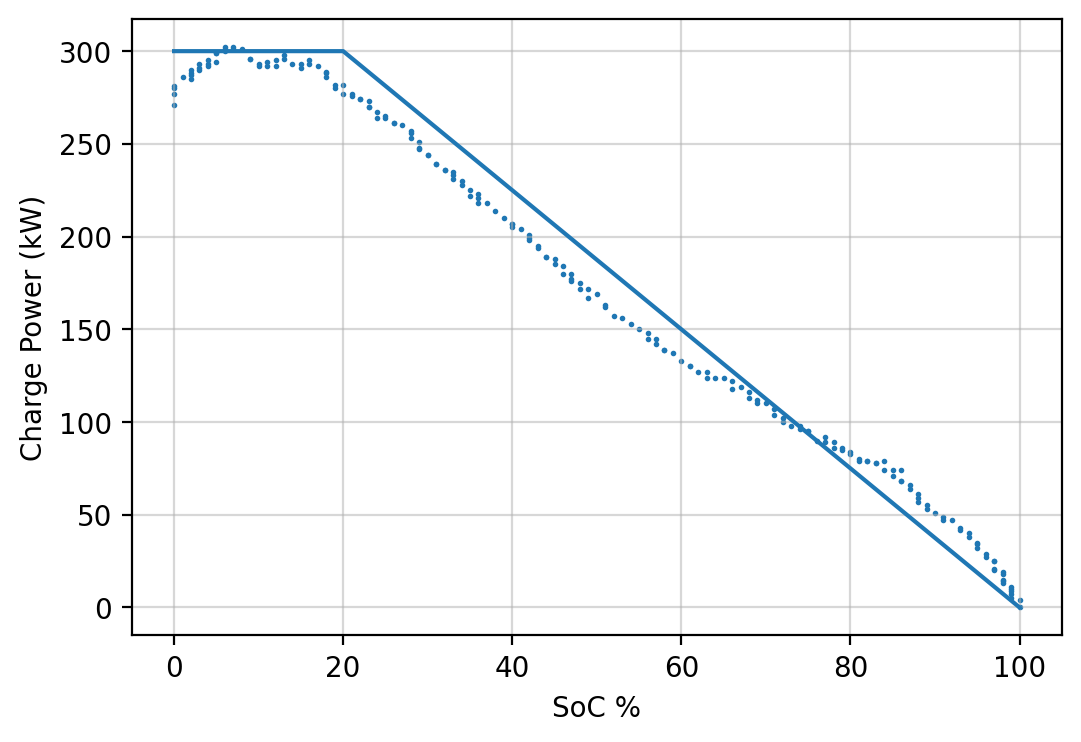}
  \caption{Experimental and simplified SoC versus charger charge power curves}\label{fig:soc_vs_chargerate}
\end{figure}

To accurately model the charging process of the eVTOLs in our simulation, we require a charge time versus State of Charge (SoC) curve. However, currently, there is no publicly available eVTOL charging time data. Therefore, to model the eVTOL charging process, we use data from an electric vehicle (EV), specifically the Lucid Air Dream Edition, which boasts the largest battery capacity and fastest charging rate among EVs to date. This vehicle has a 118 kWh battery capacity and charges from 10\% SoC to 80\% SoC in just 33 minutes using a 350 kW charger \cite{evdatabase}. Based on experimental data from the Lucid Air Dream Edition \cite{evdatabase}, we simplified the SoC - charge power relationship as shown in figure \ref{fig:soc_vs_chargerate}. In this model, we assume that the charger maintains a constant charge power up to a 20\% SoC, after which the charge power decreases linearly until the battery is fully charged. This charging behavior is modeled using a piece-wise linear function (equation \ref{SoCeq}), denoted as $f(x)$, to calculate the charge power based on the current SoC. 

\begin{equation} \label{SoCeq}
f(x)= 
\begin{cases}
M & \text{if } x \leq 20\% \text{ SoC } \\
M-S *(x-20) & \text{if } x>20\% \text{ SoC }
\end{cases}
\end{equation}

Where x is the initial SoC percentage, M is the maximum charge power and S is the slope of the line.

\begin{figure}[!ht]
  \centering
  \includegraphics[width=0.6\textwidth]{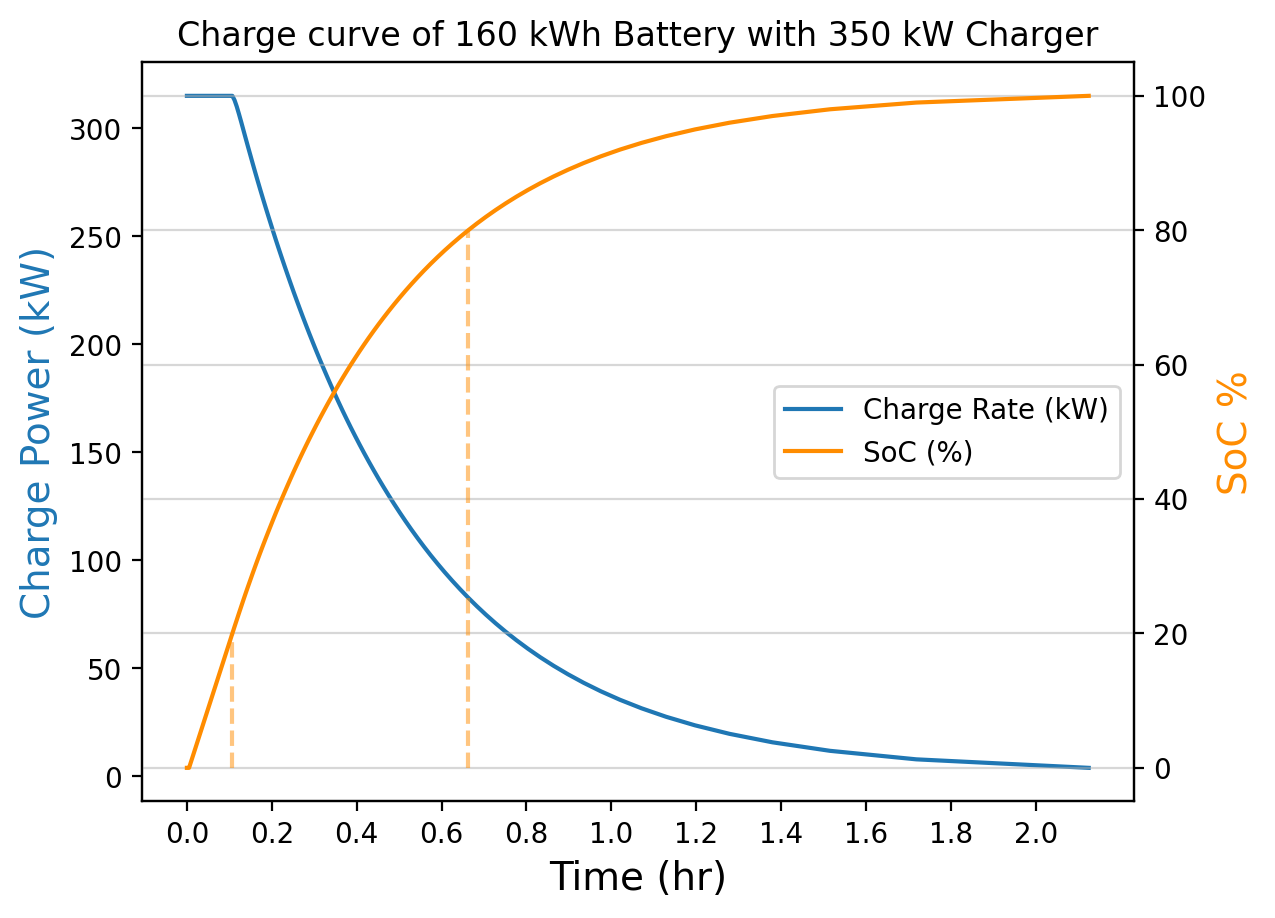}
  \caption{Time versus charge power and SoC of a battery}\label{fig:timevschargepowersoc}
\end{figure}

By integrating the charging power with respect to time, we derive the cumulative energy, as depicted in figure \ref{fig:timevschargepowersoc}. The total energy indicated by the curve in figure \ref{fig:timevschargepowersoc} should be equal to the aircraft's battery capacity, which in our case is 160 kWh.  Considering the substantial energy demand of eVTOLs, our model favors the use of 350 kW chargers. These chargers are currently the highest-powered, publicly available options for charging electric vehicles, making them a suitable choice for the high-capacity batteries used in eVTOLs \cite{terra}.

In addition, coulombic efficiency drops to \textasciitilde 90\% during DC fast charging because of the high temperatures \cite{chargerefficiency1, chargerefficiency2}. We took this energy loss into account when computing the charge power in Fig. \ref{fig:timevschargepowersoc}. 

As depicted in figure \ref{fig:timevschargepowersoc}, we can see a significant increase in charging time as the battery approaches a higher state of charge (SoC). Although charging from 0\% to 50\% takes approximately 18 minutes, charging from 0\% to 80\% requires more than twice as long, taking about 40 minutes. We also observe that charging from 20\% to 80\% takes 33 minutes. This highlights the non-linear nature of battery charging, where charging speed decreases as the battery approaches full capacity.

\subsection{Dispatch and Charge policy}
UAM operators aim to offer short travel times and on-demand flexibility akin to ride-sharing services. This introduces unique challenges for UAM, especially in traffic and fleet management given unpredictable demand. Strategic positioning of aircraft is crucial to ensure prompt response to customer requests. In this study, we follow a similar but slightly different repositioning strategy than Li et. al. \cite{Kochenderfer}. We apply the following strategies for repositioning aircraft, utilizing a similar terminology as the referenced authors:
\begin{itemize}[leftmargin=15pt]
\item Space-driven repositioning: This strategy is employed when a vertiport is at capacity, and an aircraft in the holding node is waiting to land. In such cases, any idling aircraft (those that are not charging, not embarking/disembarking passengers, and not assigned for a flight) existing at the full capacity operating vertiport are repositioned to one that has available capacity.
    \item Demand-driven repositioning: In scenarios where there is a flight request at a vertiport, but no aircraft are immediately available, the system identifies and reroutes an idling aircraft from a nearby vertiport to fulfill the request. This idling aircraft could be one that is not currently charging, not engaged in passenger embarking/disembarking, and not already assigned to a flight. In future work, we plan to investigate the charge interruption for repositioning the aircraft.
\end{itemize}

It is important to note that we do not strive for an even distribution of aircraft across vertiports. Rather, we actively reposition our fleet based on demand, taking into account the anticipated usage at different vertiports. This means we might concentrate more aircraft in high-demand areas, while leaving fewer in locations with less demand.

Regarding the charging policy in our UAM operations, we utilize insights gained from our charging model. Specifically, we commence charging an aircraft when its SoC falls to the lowest permissible limit, which we've set at a 20\% reserve SoC. Additionally, when charging an aircraft, we aim to provide enough energy for it to complete two flights, or a round trip. 


\section{Network Performance Analysis}
Our study examines the impact of fleet size and vertiport distance on performance within two-vertiport systems. We consider seven fleet sizes ranging from 8 to 20 eVTOLs, and three vertiport distances of 12, 24, and 36 miles. This analysis aims to guide stakeholders in UAM operations by illuminating the tradeoffs between passenger delay and operational costs. By understanding these metrics, eVTOL operators and vertiport developers can make informed decisions about fleet size, network design, and infrastructure development.

\subsection{Theoretical Fleet Size}
The theoretical minimum required fleet size depends on the duration of a round-trip journey, the total operating time, and the total number of flights per day on each route. The round trip time, denoted $t_{roundTrip}$, includes both the flight time and the turnaround time at each vertiport can be calculated using equation \ref{eqn:roundtrip}

\begin{equation} \label{eqn:roundtrip}
t_{roundTrip} = 2*(t_{flight} + t_{turnaround})
\end{equation}

Given the total operational time window ($t_{window}$) and the round trip time from equation \ref{eqn:roundtrip}, we can calculate the number of round trips a single aircraft can perform in 24 hours, denoted $N_{roundTrip}$, using equation \ref{eqn:num_flights}:

\begin{equation} \label{eqn:num_flights}
N_{roundTrip} = \frac{t_{window}}{t_{roundTrip}}
\end{equation}

The total number of daily flights varies for each route, denoted $d_{A-B}$ for route A-B and $d_{B-A}$ for route B-A. To satisfy the flight schedule for each route, the minimum number of aircraft required, denoted $N_{fleet}$, can be calculated using the equation \ref{eqn:fleet_size}

\begin{equation} \label{eqn:fleet_size}
N_{fleet} = \frac{max(d_{A-B}, d_{B-A})}{N_{roundTrip}}
\end{equation}

Thus, according to equation \ref{eqn:fleet_size}, the total minimum fleet size required is governed by the route with the maximum number of daily flights, as an aircraft can serve both routes. We also note that this calculation does not consider repositioning flights.

\subsection{Assumptions and Parameter Values}
\subsubsection{Assumptions}
\begin{enumerate}
    \item All nodes, both within the airspace and on the vertiport surface, have a capacity of one, with the exception of the holding node. The holding node has infinite capacity.
    \item Vertiports have one approach fix and one departure fix.
    \item A First-Reserve-First-Serve (FRFS) policy governs the allocation of all resources.
    \item Departing aircraft request access to the taxiway, TLOF, and departure fix from the parking pad before pushback.
    \item An incoming aircraft cannot start landing procedures if any aircraft has begun taxiing or has yet to release the departure fix or no parking pads are available at the vertiport.
    \item A departing aircraft cannot initiate taxiing if any other aircraft has reserved the TLOF.
\end{enumerate}

\subsubsection{Parameter Values}
\begin{itemize}
    \item Passenger waiting time threshold: 10 minutes
    \item Battery capacity: 160 kWh
    \item Charger rate: 350 kW
    \item Charge efficiency: 90 \%
    \item Number of passenger seats: 4
    \item Pre-charging processes: 3 minutes
    \item Post-charging processes: 3 minutes
    \item Taxi speed: 3.67 ft/s
    \item Embarkation and Disembarkation: Each 2 minutes
\end{itemize}

The pre-charging and post-charging processes encompass all the procedures that happen before and after the actual charging of the eVTOL battery. These may include safety checks, battery cooling and warming, and the connection and disconnection of charging infrastructure


\subsection{Fleet Size and Vertiport Configuration}
\begin{figure}[!ht]
  \centering
  \includegraphics[width=0.5\textwidth]{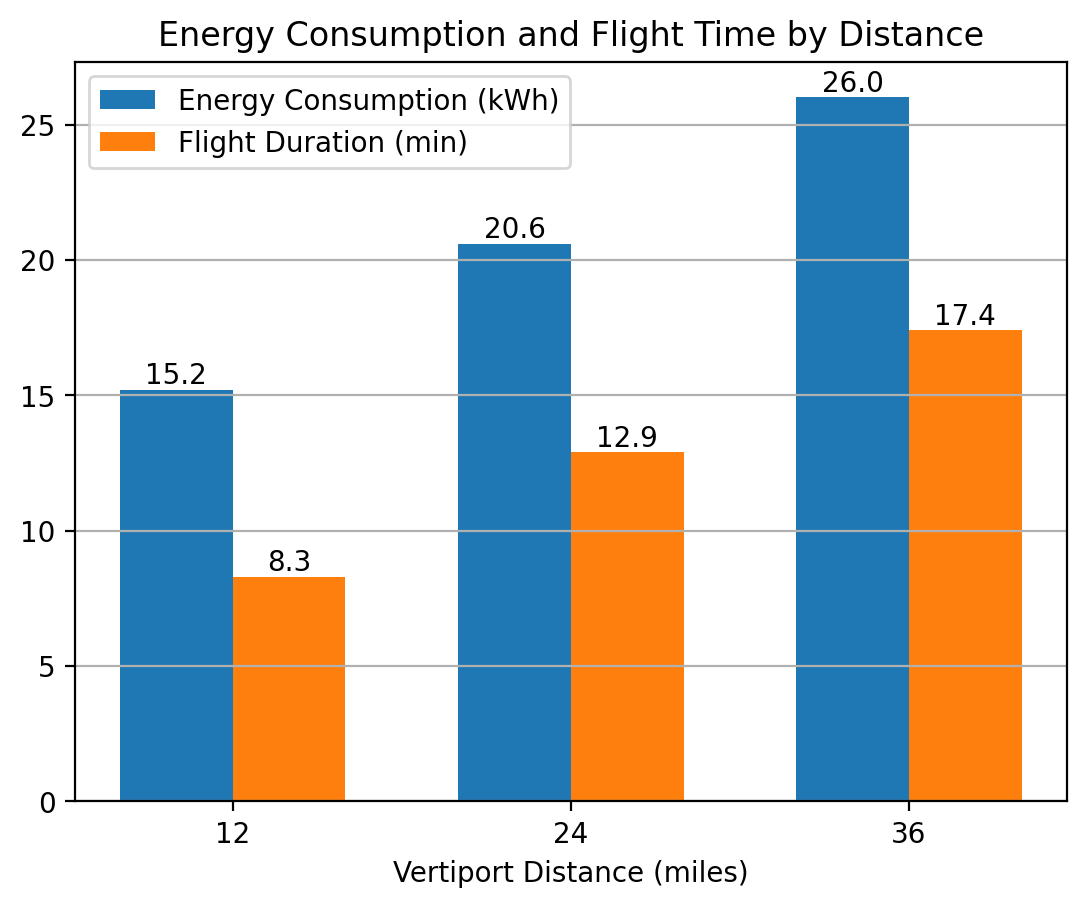}
  \caption{Energy consumption of a fully occupied aircraft and flight time between vertiports at distances of 12, 24, 36 miles}\label{fig:energy_cons_flight_time}
\end{figure}

To evaluate the minimum fleet size, we simulated three different vertiport network distances. These simulations allowed us to calculate average flight durations and energy usages, which are depicted in figure \ref{fig:energy_cons_flight_time}. With a 160 kWh battery, energy consumption led to a decrease in the SoC by 9.5\%, 12.9\%, and 16.3\% respectively for each network distance. Using these SoC decreases, we calculated necessary charging times for each round trip, resulting in turnaround times of 13, 16, and 19 minutes in the case of charging for the 12-mile, 24-mile, and 36-mile networks, respectively. Furthermore, we computed the minimum number of parking pads needed to allow for 24 departures per hour as 4, 4, and 5 for these respective turnaround times.

Utilizing equations \ref{eqn:roundtrip}, \ref{eqn:num_flights}, and \ref{eqn:fleet_size}, we find that the required minimum fleet sizes for 12-mile, 24-mile, and 36-mile networks are 11, 12, and 16 aircraft, respectively. We intentionally evaluate fleet sizes of 8, 10, 12, 14, 16, 18, and 20 to create scenarios of system underloading and overloading. In this context, "overloading" refers to deploying a fleet size smaller than the minimum requirement, potentially leading to increased passenger delays. Conversely, "underloading" denotes a fleet size larger than the minimum required, which could lead to increased operational costs, albeit with improved service times. To minimize vertiport footprints, we set the number of parking pads to half the fleet size at both vertiports. We then initiate the simulation with an equal number of aircraft at each vertiport. Similarly, to reduce the complexity of the simulation, we make the assumption that each parking pad is equipped with its own dedicated charger.

\subsection{Baseline Scenario}
\begin{figure}[!ht]
  \centering
  \includegraphics[width=0.6\textwidth]{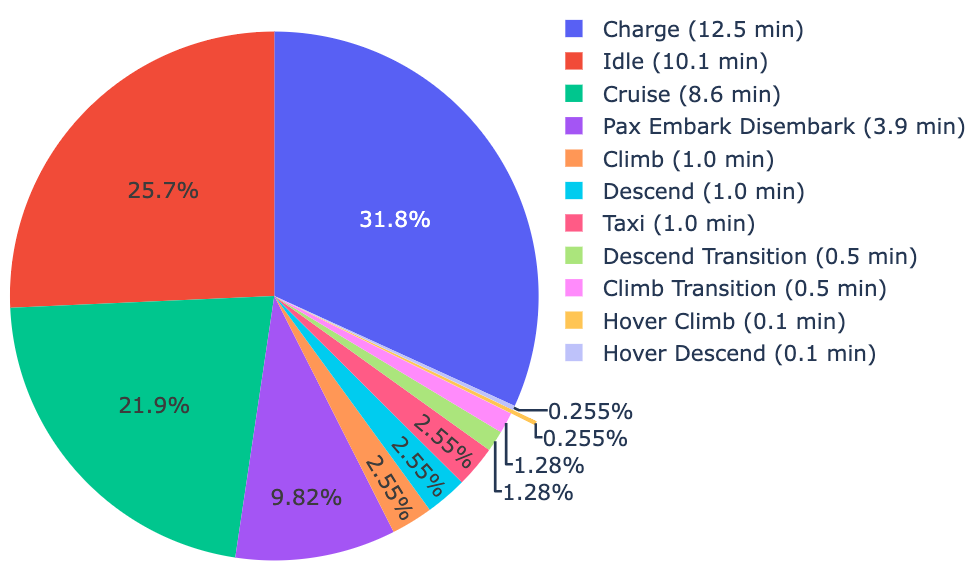}
  \caption{Average process times an aircraft experiences during a flight
}\label{fig:aircraft_lifecycle}
\end{figure}

We establish a baseline scenario with a 24-mile network and a fleet size of 14, illustrating typical flight and turnaround processes in figure \ref{fig:aircraft_lifecycle}.

Figure \ref{fig:aircraft_lifecycle} represents the average process times experienced by an aircraft during a flight for the baseline scenario. Notably, the aircraft spends the most time charging (12.5 min) and idling (10.1 min). The passenger embarkation and disembarkation take 3.9 min. The observed duration of less than 4 minutes for passenger embarkation and disembarkation processes can be attributed to certain flights that arrive and/or depart without passengers. In such cases, the process of passenger embarkation and/or disembarkation is not performed, leading to shorter operational times. These instances highlight the variability in flight load factors.

\subsection{Fleet Size, Vertiport Distance, and Passenger Delay}
\begin{figure}[!ht]
\centering
\begin{subfigure}{.5\textwidth}
  \centering
  \includegraphics[width=1\linewidth]{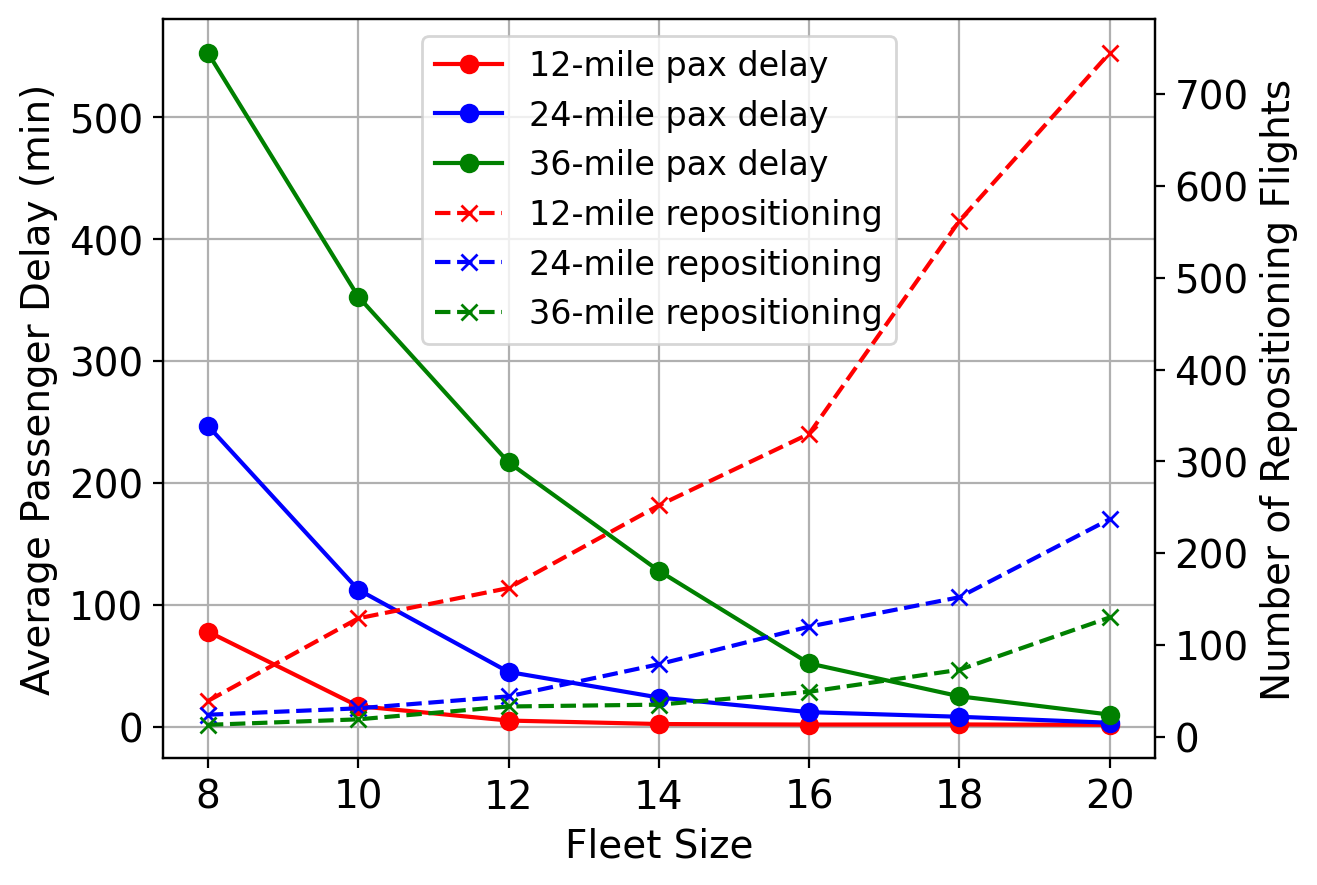}
  \caption{Fleet size effects on average passenger delays and \\repositioning flights for 12, 24, and 36-mile networks}
  \label{fig:sub1}
\end{subfigure}%
\begin{subfigure}{.5\textwidth}
  \centering
  \includegraphics[width=1\linewidth]{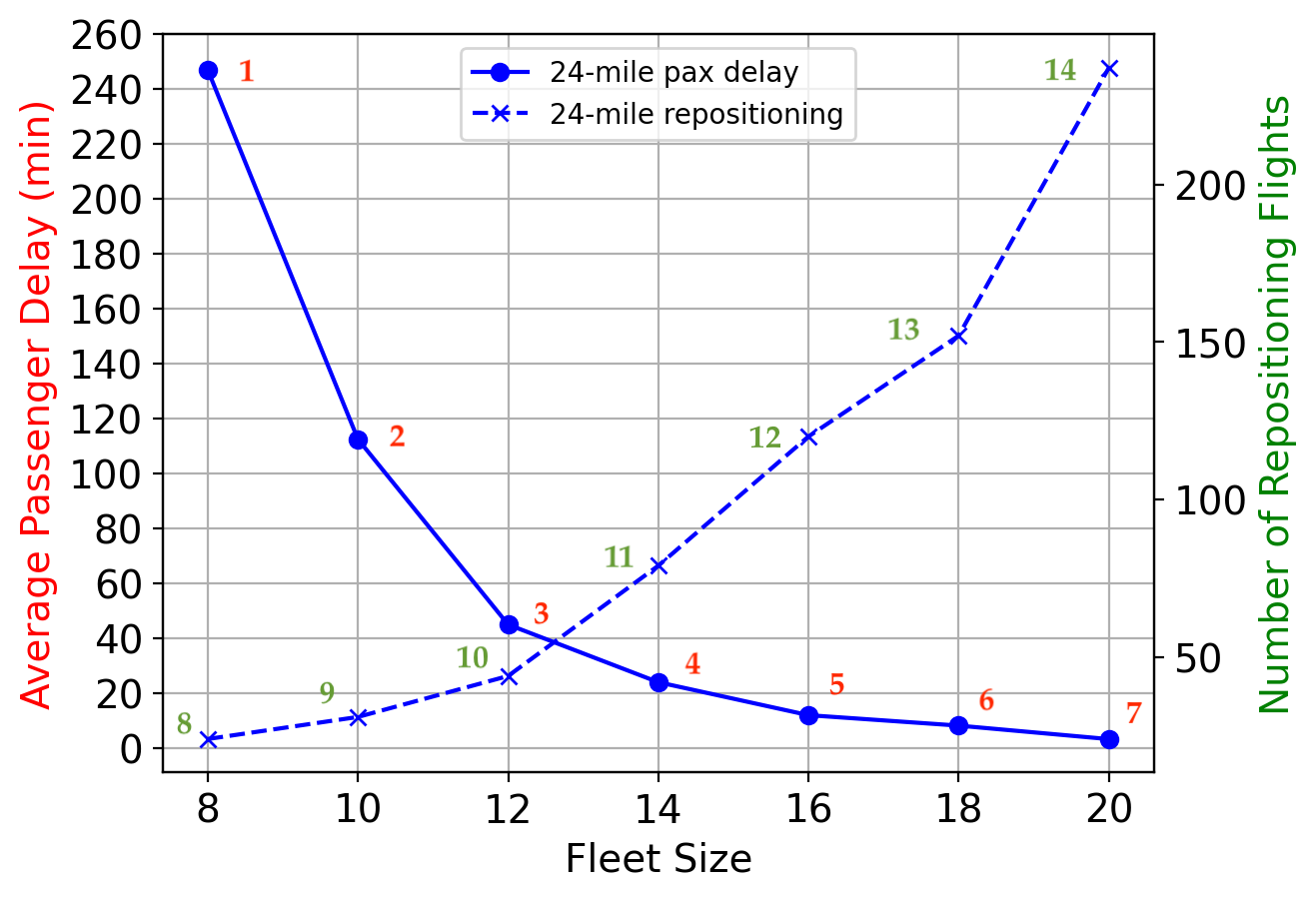}
  \caption{Focused analysis of average passenger delay versus \\repositioning flights in a 24-mile network scenario}
  \label{fig:sub2}
\end{subfigure}
\caption{Comparative evaluation of fleet size effects on average passenger delays and repositioning flights across different vertiport distances}
\label{fig:test}
\end{figure}

The primary aim of our study is to analyze the connection between fleet size, vertiport distance, and two operational metrics: average passenger delay and the number of repositioning flights. Figure \ref{fig:sub1} illustrates the relationship between average passenger delay and fleet size, represented by solid lines. It can be observed across all vertiport distances that a larger fleet size reduces the average delay. Nevertheless, this benefit comes with a significant trade-off: an escalated number of repositioning flights. This effect is primarily due to an excess of available aircraft coupled with the limited capacity at vertiports. To accommodate incoming aircraft and meet the demand, both space-driven and demand-driven repositioning policies assign more aircraft for repositioning. In every scenario we considered, the number of parking pads is equal to the number of aircraft.

Our study found that a 12-mile distance could meet the maximum waiting time threshold of 10 minutes with a fleet of 12 aircraft, yielding an average passenger delay of 5.1 minutes. For a 24-mile (network), 18 aircraft were needed (resulting in an average passenger delay of 8.2 minutes), while a 36-mile (network) required 20 aircraft (causing an average passenger delay of 10 minutes). These fleet sizes are larger than those calculated using equation \ref{eqn:fleet_size}. The reason being that equation \ref{eqn:fleet_size} determines fleet size under balanced demand without considering repositioning flights. Hence, it tends to underestimate the actual fleet size in instances of imbalanced demand.

Another interesting observation is that the 12-mile system could accommodate a surplus of fleet size up to a specific limit (an additional 7 aircraft than that computed with equation \ref{eqn:fleet_size}) before the number of repositioning flights rose significantly. Meanwhile, the longer flight and turnaround times associated with the 24-mile network could absorb a sharp increase in repositioning flights up to a fleet size of 20. Further increase in the fleet size triggered a sharp uptick in repositioning flights, mirroring the pattern observed in the 12-mile scenario. This data is not depicted in the figure for readability purposes.

For a more detailed understanding, refer to figure \ref{fig:sub2}, which maps the trade-off between a decreasing average passenger delay and an increasing number of repositioning flights. In this graph, reducing the delay from 247 to 112 minutes (from point 1 to point 2) necessitates 7 additional repositioning flights (from 24 to 31) and adding two more aircraft. To further cut down the delay to 45 minutes, the cost is an additional 13 repositioning flights (from 31 to 44) and two more aircraft. Beyond this point, although the passenger delay continues to decrease, the rate of this decrease slows down. Simultaneously, the rate of increase in repositioning flights intensifies. Finally, in an 18-aircraft system, the average passenger delay can be maintained at 8.2 minutes, but this results in a substantial increase in the number of repositioning flights, escalating to a total of 152. 

A viable strategy to reduce the number of repositioning flights involves increasing the number of parking pads. However, this approach entails higher capital expenditure for additional infrastructure. The decision between incurring larger infrastructure costs or bearing the expense of repositioning flights is a complex consideration that will be explored in a future study.

\begin{figure}[!ht]
  \centering
  \includegraphics[width=0.5\textwidth]{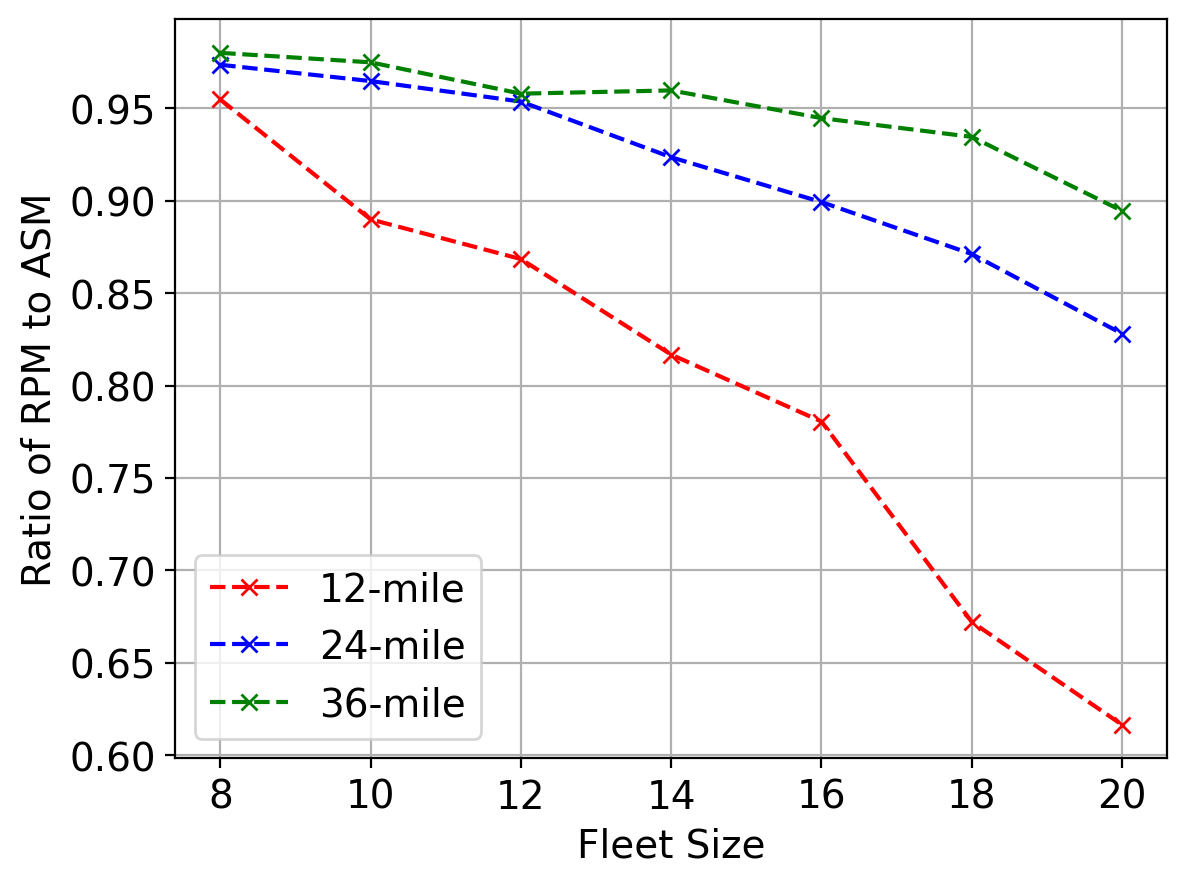}
  \caption{Ratio of revenue passenger miles to available seat miles as a function of fleet size}\label{fig:rpm_asm}
\end{figure}

In figure \ref{fig:rpm_asm} we present the ratio of Revenue Passenger Miles (RPM) to Available Seat Miles (ASM). RPM and ASM are essential metrics in the airline industry. RPM measures an airline's passenger traffic, calculated by multiplying the number of paying passengers by the distance they travel. On the other hand, ASM represents an airline's flight capacity, derived by multiplying the total number of seats available on all flights by the distance of those flights \cite{Fielding1978}. It is important to note that in our simulations, the repositioning flights may carry passengers. This is because the \textit{System Manager} consistently checks the waiting rooms prior to dispatching a flight in an effort to maximize load factors. Despite this, these flights are still classified as repositioning flights due to their primary purpose of repositioning the aircraft. This factor is incorporated in the calculation of RPM.

\subsection{Fleet Utilization and Energy Consumption Analysis}
To analyze the impact of vertiport distances on fleet utilization, we conducted simulations across three scenarios: 12-mile, 24-mile, and 36-mile vertiport distances. Table \ref{tab:vertiport_distance_util} presents the daily total values of various fleet utilization metrics, both for the whole network and per individual aircraft. For all the scenarios, we set the fleet size to 14 aircraft. Our findings show a clear correlation between vertiport distances and aircraft utilization. As the vertiport distance increases, the total idle time across the network decreases, while the total cruise time and charge time increase. Concurrently, there is a reduction in the number of repositioning flights, contributing to the decline in the total number of flights, marking a more efficient utilization of the fleet. However, these enhancements in fleet utilization come with trade-offs. Notably, the average passenger delay increases from 2.3 minutes in the 12-mile network to 24 minutes in the 24-mile network, and drastically to 128 minutes in the 36-mile network. The increased delay significantly exceeds our acceptable threshold of 10 minutes in the 36-mile network scenario. Thus, while the 36-mile network appears to be the most efficient in terms of fleet utilization, the resultant passenger delay might be unacceptable to many passengers.

Additionally, as expected, as vertiport distance extends, total charging time rises, peaking at 36 miles. The longer charging times results in less availability for flight. Additionally, holding times decline with increased vertiport distance due to the additional capacity at vertiports from having more aircraft airborne concurrently. The 'Other' category includes processes, aside from charging, idling, cruising, and holding, as presented in Figure \ref{fig:aircraft_lifecycle}. Given that the number of flights decreased and the rate of change for idle, charge, and cruise processes was higher than that for 'Other' processes, we observed a corresponding decrease in the percentage of 'Other' processes as vertiport distance increased.

\begin{table}[!ht]
\caption{Fleet utilization with respect to varying vertiport distances}
\label{tab:vertiport_distance_util}
\centering
\begin{tabular}{|c|cc|cc|cc|}
\hline
\begin{tabular}[c]{@{}c@{}}Vertiport\\ Distance\end{tabular}   & \multicolumn{2}{c|}{12 miles}                                                                                                                              & \multicolumn{2}{c|}{24 miles}                                                                                                                              & \multicolumn{2}{c|}{36 miles}                                                                                                                               \\ \hline
\begin{tabular}[c]{@{}c@{}}Aircraft\\ Utilization\end{tabular} & \multicolumn{1}{c|}{\begin{tabular}[c]{@{}c@{}}Network\\ Total (hrs)\end{tabular}} & \begin{tabular}[c]{@{}c@{}}Average Per \\ Aircraft (hrs)\end{tabular} & \multicolumn{1}{c|}{\begin{tabular}[c]{@{}c@{}}Network\\ Total (hrs)\end{tabular}} & \begin{tabular}[c]{@{}c@{}}Average Per \\ Aircraft (hrs)\end{tabular} & \multicolumn{1}{c|}{\begin{tabular}[c]{@{}c@{}}Network\\ Total (hrs)\end{tabular}} & \begin{tabular}[c]{@{}c@{}}Average Per  \\ Aircraft (hrs)\end{tabular} \\ \hline
Idle                                                           & \multicolumn{1}{c|}{145.43}                                                        & 10.39                                                                  & \multicolumn{1}{c|}{97.29}                                                         & 6.95                                                                  & \multicolumn{1}{c|}{68.33}                                                         & 4.88                                                                   \\ \hline
Charge                                                         & \multicolumn{1}{c|}{72.01}                                                         & 5.14                                                                  & \multicolumn{1}{c|}{71.56}                                                         & 5.11                                                                  & \multicolumn{1}{c|}{111.25}                                                         & 7.95                                                                   \\ \hline
Cruise                                                         & \multicolumn{1}{c|}{50.89}                                                         & 3.64                                                                  & \multicolumn{1}{c|}{101.03}                                                         & 7.22                                                                  & \multicolumn{1}{c|}{146.81}                                                        & 10.49                                                                   \\ \hline
Holding                                                        & \multicolumn{1}{c|}{8.51}                                                          & 0.61                                                                  & \multicolumn{1}{c|}{4.47}                                                          & 0.32                                                                  & \multicolumn{1}{c|}{2.14}                                                          & 0.15                                                                   \\ \hline 
Other                                                          & \multicolumn{1}{c|}{76.61}                                                         & 5.47                                                                  & \multicolumn{1}{c|}{73.58}                                                         & 5.26                                                                  & \multicolumn{1}{c|}{50.89}                                                         & 3.64                                                                   \\ \hline \hline
\# Flights  &
\multicolumn{1}{c|}{867}                                                         & 62                                                                  & \multicolumn{1}{c|}{765}                                                         & 55                                                                  & \multicolumn{1}{c|}{735}                                                         & 53                                                                   \\ \hline
\end{tabular}
\end{table}

\begin{figure}[!ht]
  \centering
  \includegraphics[width=1\textwidth]{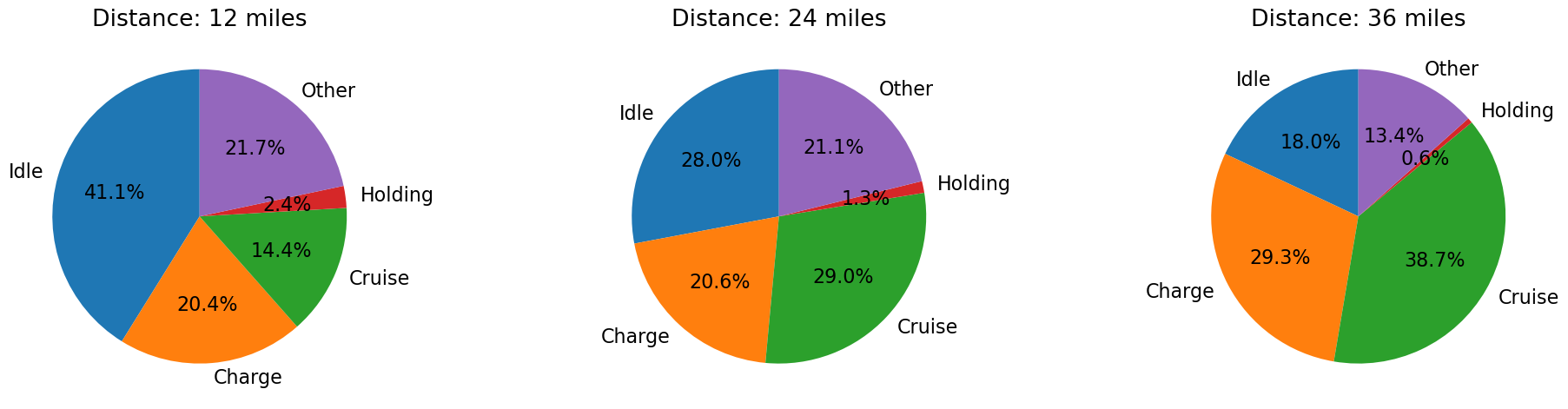}
  \caption{Fleet utilization for vertiport networks at distances of 12, 24, and 36 miles for the baseline fleet size of 14. Pie chart created using the total network hours}\label{fig:distance_aircraft_util}
\end{figure}

We also analyze fleet utilization across varying fleet sizes for a fixed 24-mile vertiport network. To do this, we considered fleet sizes of 8, 12, 16, and 20, respectively, as shown in Table \ref{tab:fleet_util_fleet}. The table summarizes the total network and average per aircraft utilization metrics for each fleet size. We observe that as the fleet size increases, the total number of flights across the network also increases due to increased repositioning flights. However, despite the overall increase in the number of flights, the average number of flights per aircraft decreases. This can be interpreted as each aircraft undertaking fewer individual flights, as the larger fleet size allows the flight load to be distributed across more aircraft. We see a similar trend in the charge time as well.

Similarly, increasing fleet size notably raises total idle time, as flight loads are distributed among more aircraft. This increase, while seemingly inefficient, actually provides a buffer for unexpected events or fluctuating demand. Conversely, the average passenger delay decreases as the fleet size expands (with average passenger delays of 247 mins, 45 mins, 12 mins, and 3.28 mins for fleet sizes of 8, 12, 16, and 20, respectively). The challenge lies in balancing fleet size to minimize both passenger delay and idle time, optimizing for operational efficiency and passenger experience.

\begin{table}[!ht]
\caption{Fleet utilization with respect to varying fleet sizes for the baseline network}
\label{tab:fleet_util_fleet}
\centering
\begin{tabular}{|c|cc|cc|cc|cc|}
\hline
Fleet Size                                                     & \multicolumn{2}{c|}{8}                                                                                                                                              & \multicolumn{2}{c|}{12}                                                                                                                                             & \multicolumn{2}{c|}{16}                                                                                                                                              & \multicolumn{2}{c|}{20}                                                                                                                                             \\ \hline
\begin{tabular}[c]{@{}c@{}}Aircraft\\ Utilization \\ (hrs)\end{tabular} & \multicolumn{1}{c|}{\begin{tabular}[c]{@{}c@{}}Network\\ Total\end{tabular}} & \begin{tabular}[c]{@{}c@{}}Average \\ Per \\ Aircraft\end{tabular} & \multicolumn{1}{c|}{\begin{tabular}[c]{@{}c@{}}Network\\ Total\end{tabular}} & \begin{tabular}[c]{@{}c@{}}Average \\ Per \\ Aircraft\end{tabular} & \multicolumn{1}{c|}{\begin{tabular}[c]{@{}c@{}}Network\\ Total\end{tabular}} & \begin{tabular}[c]{@{}c@{}}Average \\ Per  \\ Aircraft\end{tabular} & \multicolumn{1}{c|}{\begin{tabular}[c]{@{}c@{}}Network\\ Total\end{tabular}} & \begin{tabular}[c]{@{}c@{}}Average \\ Per \\ Aircraft\end{tabular} \\ \hline
Idle                                                           & \multicolumn{1}{c|}{39.69}                                                            & 4.96                                                                        & \multicolumn{1}{c|}{79.39}                                                            & 6.62                                                                        & \multicolumn{1}{c|}{132.60}                                                           & 8.29                                                                         & \multicolumn{1}{c|}{197.12}                                                           & 9.86                                                                       \\ \hline
Charge                                                         & \multicolumn{1}{c|}{66.49}                                                            & 8.31                                                                        & \multicolumn{1}{c|}{66.90}                                                            & 5.58                                                                        & \multicolumn{1}{c|}{75.64}                                                            & 4.73                                                                         & \multicolumn{1}{c|}{81.09}                                                            & 4.05                                                                        \\ \hline
Cruise                                                         & \multicolumn{1}{c|}{92.46}                                                            & 11.56                                                                        & \multicolumn{1}{c|}{95.23}                                                            & 7.94                                                                        & \multicolumn{1}{c|}{105.32}                                                           & 6.58                                                                         & \multicolumn{1}{c|}{110.87}                                                           & 5.54                                                                        \\ \hline
Holding                                                        & \multicolumn{1}{c|}{1.55}                                                             & 0.19                                                                        & \multicolumn{1}{c|}{2.74}                                                             & 0.23                                                                        & \multicolumn{1}{c|}{6.22}                                                             & 0.39                                                                         & \multicolumn{1}{c|}{9.18}                                                             & 0.46                                                                        \\ \hline
Other                                                          & \multicolumn{1}{c|}{67.58}                                                            & 8.45                                                                        & \multicolumn{1}{c|}{69.97}                                                            & 5.83                                                                        & \multicolumn{1}{c|}{76.03}                                                            & 4.75                                                                         & \multicolumn{1}{c|}{78.93}                                                            & 3.95                                                                        \\ \hline \hline
\# Flights  &
\multicolumn{1}{c|}{727}                                                         & 91                                                                  & \multicolumn{1}{c|}{741}                                                         & 62                                                                  & \multicolumn{1}{c|}{787}                                                         & 49             
                & \multicolumn{1}{c|}{853}                                                         & 43       
\\ \hline
\end{tabular}
\label{tab:fleetsize_fleet_util}
\end{table}

\begin{figure}[!ht]
  \centering
  \includegraphics[width=1\textwidth]{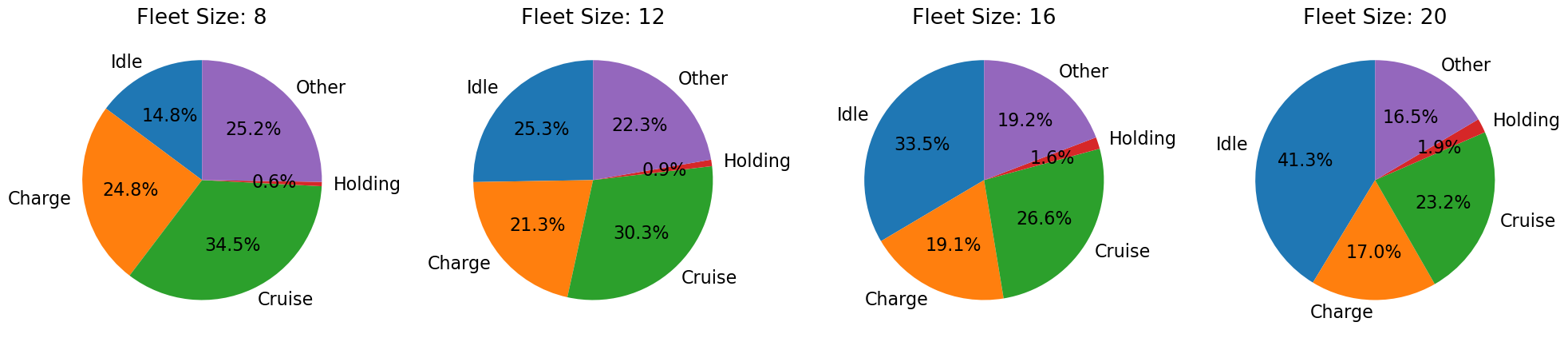}
  \caption{Fleet utilization for the baseline network of 24-mile. Pie chart created using the total network hours}\label{fig:fleetsize_aircraft_util}
\end{figure}

\begin{figure}[!ht]
  \centering
  \includegraphics[width=0.8\textwidth]{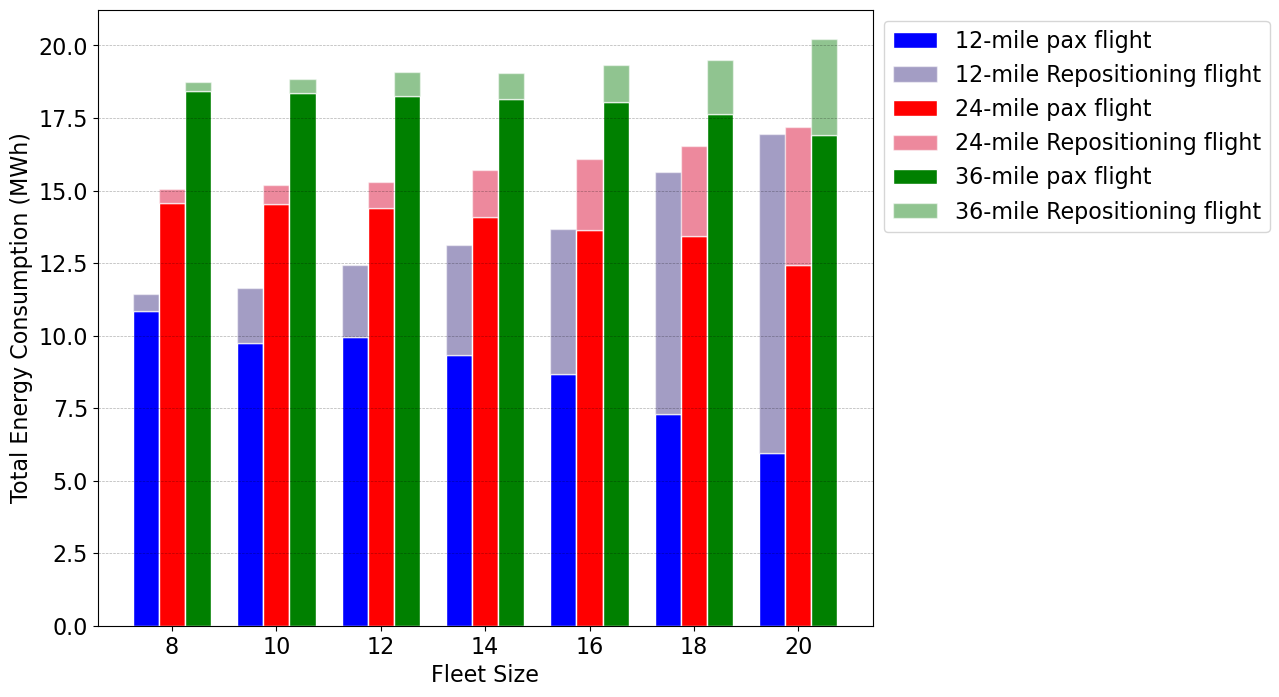}
  \caption{Total daily energy consumption of the network as a function of fleet size. The darker bars represent passenger flights, while the lighter bars represent repositioning flights}\label{fig:energy_cons_fleet_size}
\end{figure}

The analysis of energy consumption across varying fleet sizes and vertiport distances shown in figure \ref{fig:energy_cons_fleet_size} supports the previous results. As fleet size increases, the network's total energy consumption rises due to the increase in repositioning flights, even though the energy consumed by the passenger flight decreases. This points to a trade-off between energy efficiency and operational efficiency in managing the fleet size for networks of 12-mile, 24-mile, and 36-mile distances. 

\section{Conclusion and Future Work}
Our analysis with VertiSim provides a better understanding of the operational complexities and tradeoffs involved in UAM systems. By studying a range of scenarios with different fleet sizes and vertiport distances, we were able to quantify the impact of these variables on performance metrics such as passenger delay time, energy consumption, and fleet utilization.  This research could offer useful insights to e-VTOL operators and vertiport developers in optimizing their operational strategies and infrastructure development. Our study highlights key operational differences between UAM and legacy aviation. A pivotal finding is the substantial role of charging times in UAM, consuming over a quarter of an aircraft's operational time and significantly affecting fleet utilization. We also observed that an increase in fleet size leads to more repositioning flights. This trend is partly due to our simulation setup, where the total number of parking pads in the system was kept equal to the number of aircraft, leading to increased repositioning to manage aircraft availability and vertiport occupancy. In essence, we have underscored the importance of striking a balance between passenger service quality and operational efficiency for the successful deployment and sustainability of future UAM systems. Our hope is that VertiSim can continue to be a valuable tool for investigating the multifaceted challenges of UAM operations. 

Looking forward, the application of VertiSim to more complex vertiport networks and scheduled flights offers an exciting avenue for exploration. Such flights necessitate a distinct set of considerations for optimizing dispatch and charging policies. To address these nuances, our future work will aim to incorporate a system-level optimization approach, integrating these policies within the simulation framework. Additionally, we plan to incorporate wind and weather conditions into our simulations. This inclusion will allow us to simulate the impact of environmental factors on flight durations, energy consumption, and vertiport operations, thereby enhancing the realism and utility of VertiSim in planning and managing urban air mobility networks.

\bibliography{references}

\end{document}